\documentclass[aps,prl,showpacs,amsmath,amssymb,amsfonts,twocolumn, superscriptaddress,floatfix,footinbib]{revtex4-1}
\usepackage[T1]{fontenc}
\usepackage{graphicx}
\usepackage{color}
\usepackage{ulem}
\usepackage{hyperref}

\newlength{\halfcolumnwidth}
\begin{document}
\title{Direct Observation of Dirac Cones and a Flatband in a Honeycomb Lattice for Polaritons}

\date{\today}
\author{T.~Jacqmin}
\affiliation{Laboratoire de Photonique et Nanostructures, CNRS/LPN, Route de Nozay, 91460 Marcoussis, France}

\author{I.~Carusotto}
\affiliation{INO-CNR BEC Center and Dipartimento di Fisica, Universit\`a di Trento, I-38123, Povo, Italy}

\author{I.~Sagnes}
\affiliation{Laboratoire de Photonique et Nanostructures, CNRS/LPN, Route de Nozay, 91460 Marcoussis, France}

\author{M.~Abbarchi}
\altaffiliation{Present address, IM2NP, Aix-Marseille Université, UMR CNRS 6242, 13997 Marseille, France}
\affiliation{Laboratoire de Photonique et Nanostructures, CNRS/LPN, Route de Nozay, 91460 Marcoussis, France}

\author{D.~D.~Solnyshkov}
\affiliation{Institut Pascal, PHOTON-N2, Clermont Université, Université Blaise Pascal, CNRS, 24 avenue des Landais, 63177 Aubière Cedex, France}

\author{G.~Malpuech}
\affiliation{Institut Pascal, PHOTON-N2, Clermont Université, Université Blaise Pascal, CNRS, 24 avenue des Landais, 63177 Aubière Cedex, France}

\author{E.~Galopin}
\affiliation{Laboratoire de Photonique et Nanostructures, CNRS/LPN, Route de Nozay, 91460 Marcoussis, France}

\author{A.~Lema\^itre}
\affiliation{Laboratoire de Photonique et Nanostructures, CNRS/LPN, Route de Nozay, 91460 Marcoussis, France}

\author{J.~Bloch}
\affiliation{Laboratoire de Photonique et Nanostructures, CNRS/LPN, Route de Nozay, 91460 Marcoussis, France}

\author{A.~Amo}
\affiliation{Laboratoire de Photonique et Nanostructures, CNRS/LPN, Route de Nozay, 91460 Marcoussis, France}

\begin{abstract}
Two-dimensional lattices of coupled micropillars etched in a planar semiconductor microcavity offer a workbench to engineer the band structure of polaritons. We report experimental studies of honeycomb
lattices where the polariton low-energy dispersion is analogous to that of electrons in graphene. Using energy-resolved photoluminescence we directly observe Dirac cones, around which the dynamics of
polaritons is described by the Dirac equation for massless particles. At higher energies, we observe p orbital bands, one of them with the non-dispersive character of a flatband. The realization of this
structure which holds massless, massive and infinitely-massive particles opens the route towards studies of the interplay of dispersion, interactions, and frustration in a novel and controlled environment.
\end{abstract}


\maketitle 

Engineering Hamiltonians in controlled systems has proven to be a useful tool to simulate and unveil complex condensed matter phenomena otherwise experimentally inaccessible. Indeed, condensed-matter systems usually lack control and observables, whereas model systems such as ultracold atoms~\cite{Bloch2008}, arrays of photonic waveguides~\cite{Szameit2010}, or polariton gases~\cite{Carusotto2013} enable the control of the density, the temperature, and in the case of lattice systems, the topology of the band structure. 
In this context, the honeycomb lattice, whose geometry is responsible for
the properties of graphene. has attracted a lot of attention. This extraordinary material shows pointlike intersections between the conduction and valence bands. Around those points, referred to as Dirac points, the energy dispersion is linear, and electrons behave like massless relativistic particles~\cite{CastroNeto2009}. The honeycomb geometry gives rise to intriguing phenomena such as anomalous Klein tunneling and geometric phase effects that result in the antilocalization of electrons~\cite{CastroNeto2009}. In addition, geometric frustration in the honeycomb lattice is expected to give rise to nondispersive
bands in which all states are localized~\cite{Wu2007}. These bands have not yet been experimentally evidenced.

The investigation of this physics has triggered the realization of simulators~\cite{Polini2013} whose parameters can be controlled in a range not easily accessible in graphene. 
For instance, honeycomb lattices for cold atoms~\cite{Soltan-Panahi2011, Tarruell2012}, electrons gases in solids~\cite{DeSimoni2010} and molecules~\cite{Gomes2012} and acoustic waves~\cite{Torrent2012} were realized. In photonics,  honeycomb lattices were created using light-induced lattices in nonlinear crystals~\cite{Peleg2007}, microwave-domain photonic crystals~\cite{Bittner2010}, arrays of coupled waveguides~\cite{Rechtsman2012b, Rechtsman2013}, and resonators~\cite{Bellec2013}. While these systems have shown remarkable features like topological phase transitions~\cite{Tarruell2012} or the possibility of including synthetic gauge fields~\cite{Rechtsman2012b}, they lack simultaneous control of the particle momentum, local potential, interactions and on-site visualization. In this sense, polaritons in semiconductor planar microcavities appear as an extraordinary platform overcoming these limitations~\cite{Carusotto2013}. These light-matter particles, which arise from the strong coupling between cavity photons and quantum well excitons can be created, manipulated and detected using optical techniques. Two-dimensional lattices for polaritons have been implemented using surface acoustic waves~\cite{Cerda-Mendez2012a} and gold deposition at the surface of the cavity~\cite{Kim2011, Kim2013, Kusudo2013}. However, the former method allows very limited lattice geometries, while the latter can only provide very shallow modulations of the potential. Alternatively, the recent realization of coupled micropillars based on deep etching of a planar structure~\cite{Dousse2010, Galbiati2012} has opened the way towards the engineering of lattices for polaritons with controlled tunneling and deep on-site potentials with arbitrary geometry.

In this Letter, we report on a honeycomb lattice for polaritons, made of hundreds of coupled micropillars etched in a planar semiconductor microcavity. By monitoring the photoluminescence at low excitation density, we directly image the energy dispersion of the structure, which reveals several energy bands. The lowest two arise from the coupling between the fundamental modes of the micropillars. They are analogous to the $\pi$ and $\pi^*$ bands of graphene~\cite{CastroNeto2009}. In particular, we evidence six Dirac cones at the corners of the first Brillouin zone (Bz), around which the energy dispersion is linear. When increasing the excitation intensity, we observe polariton condensation occurring at the top of the $\pi^*$ band, showing spatial coherence extended over the whole excitation spot.
Additionally, we report on the presence of higher-energy bands arising from the coupling between higher-energy modes of the pillars. In particular, we observe a nondispersive band in which polaritons have an infinite effective mass. The observation of this flatband opens the way to the study  of the interplay of interactions, frustration, and spin dynamics in a
novel driven-dissipative framework.

\begin{figure}
\includegraphics[scale=0.55]{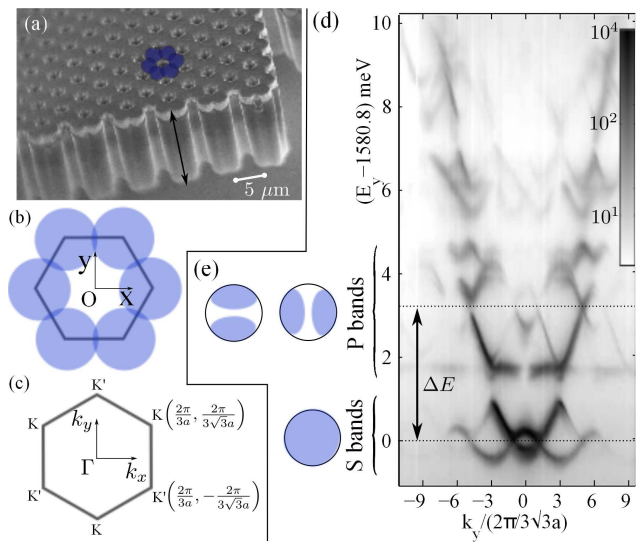}
\caption{a) Scanning electron microscope image of a corner of the microstructure. One hexagon of pillars is underlined with blue disks. The dark arrows show the growth axis of the cavity. The overlap between pillars is sketched in (b). (c) First Bz. (d) Measured momentum space energy resolved photoluminescence at $k_x=-2\pi/3a$ [line 0 in Fig.~\ref{fig2}(a)], under nonresonant low-power excitation. (e) Sketch of the real space distribution of $S$ and $P$ modes in a single pillar.}
\label{fig1}
\end{figure}

Our structure is a $Q = 72 000$ $ \lambda /2$ microcavity. It is a $\rm{Ga}_{0.05}\rm{Al}_{0.95}\rm{As}$ layer surrounded by two $\rm{Ga}_{0.05}\rm{Al}_{0.95}\rm{As}/\rm{Ga}_{0.8}\rm{Al}_{0.2}\rm{As}$ Bragg mirrors with 28 (40) top (bottom) pairs respectively. Twelve GaAs quantum wells of 7~nm width are inserted inside the cavity, yielding a 15~meV Rabi splitting. Experiments are performed at 10~K and -17~meV cavity-exciton detuning. We engineer a honeycomb lattice of coupled micropillars by using electron beam lithography and dry etching of the sample down to the GaAs substrate [see Fig.~\ref{fig1}(a)]. The diameter of each pillar is $d=3~\mu$m, and the distance between two adjacent pillars (the lattice constant), is $a = 2.4~\mu$m. 
The etched cavity shows polariton lifetime of 27~ps at the bottom of the lower polariton band.
As the interpillar distance is smaller than their diameter, the pillars spatially overlap [see Fig.~\ref{fig1}(b)]. This results in a sizable polariton tunnel coupling between adjacent micropillars via their photonic component~\cite{Galbiati2012}. For our structure, the tunnel coupling amounts to 0.25~meV. 
The system is excited out of resonance with a Ti:Sapph monomode laser at 730 nm, in a spot of 30~$\mu$m diameter covering around 30 pillars. The photoluminescence is collected through a high numerical aperture objective (NA = 0.65), dispersed in a spectrometer and detected by a CCD camera on which we can image either the real or the momentum space. Note that a chopper was used in the case of high power excitation to avoid heating of the sample.

Under low-power excitation, incoherent relaxation of polaritons results in the population of all the energy bands. Note that for low power excitation polariton-polariton interactions are negligible so that single particle physics of the honeycomb lattice is probed. Figure~\ref{fig1}(d) shows the measured far field photoluminescence containing many groups of bands, separated by energy gaps. 
The two lowest bands ($S$ bands) arise from the coupling between the fundamental mode of the pillars ($S$ modes). At higher energy, we observe a group of four bands ($P$ bands) arising from the coupling between the first excited state of the pillars, which is twice degenerate and has two lobes~\cite{Galbiati2012} [see Fig.~\ref{fig1}(e)]. The separation between these two groups of bands is $\Delta E = 3.2~$meV, the energy difference between the two lowest-energy states of the individual pillars. Above those two groups of bands, many others can be seen arising from the hybridization of higher energy modes of the pillars.

\begin{figure}
\includegraphics[scale=1.18]{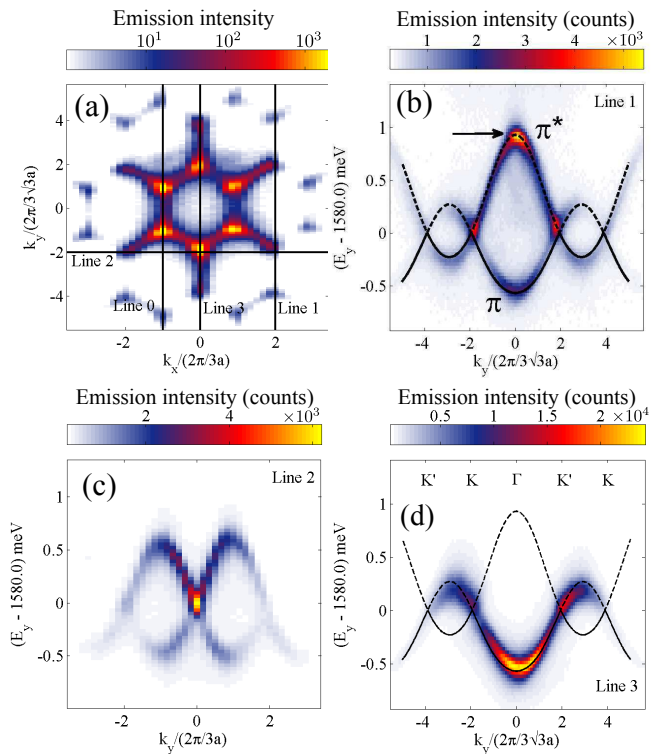}
\caption{(a) Measured photoluminescence intensity in momentum space at the energy of the Dirac points [dotted line in Fig.\ref{fig1}(d)]. (b) Spectrally resolved far field emission along line~1 in (a). The black line is a fit to Eq.~(\ref{tb_dispersion}). (c) Same as (b) along line~2 in (a). (d) Spectrally resolved far field emission along line 3 in (a), passing through the first Bz.}
\label{fig2}
\end{figure}

The two $S$ bands stem from the coupling between micropillar states which have a cylindrical symmetry similar to that of the carbon $P_z$ electronic orbitals in graphene. Thus, we expect the two $S$ bands to present features analogous to the $\pi$ and $\pi^*$ bands of graphene, including six Dirac (contact) points~\cite{CastroNeto2009} in the first Bz [see Fig.~\ref{fig1}(c)]. Figure~\ref{fig2}(a) shows the measured emitted intensity in momentum space at the Dirac points energy [zero energy in Fig.~\ref{fig1}(d)]. We observe the six Dirac points at the corner of the first Bz (yellow points). The adjacent Bzs are also seen. Figures~\ref{fig2}(b and c) show the measured energy resolved emission along the lines 1 and 2 indicated in Fig.~\ref{fig2}(a), passing through four and three Dirac linear intersections respectively.
As the confinement energy on each site of the lattice is much larger than the tunneling energy, the system is well described by the tight-binding approximation. Including first- and second-neighbor tunneling the following dispersion can be obtained~\cite{CastroNeto2009},
\begin{equation}
E(\textbf{k}) = \pm t \sqrt{3+f(\textbf{k})}-t^{'}f(\textbf{k}),
\label{tb_dispersion}
\end{equation}
where
\begin{equation}
f(\textbf{k}) = 2\cos{\left(\sqrt{3}k_y a\right)}+4\cos{\!\left(\frac{\sqrt{3}}{2}k_y a\right)\!}\cos{\!\left(\frac{3}{2}k_x a\right)\!}.
\end{equation}
By fitting Eq.~(\ref{tb_dispersion}) to the data in Fig.~\ref{fig2} we extract a value of the coupling between first and second neighbors of $t = 0.25$ and $t^{'}=-0.02$~meV, respectively. The result of the fit is shown in Fig.~\ref{fig2}(b), and yields a group velocity $v = 3at/2\hbar = 1.3\times 10^{6}$~m.s$^{-1}$ around the Dirac points. Note that the data shown in Fig.~\ref{fig2}(b) do not belong to the first Bz. If we perform the same measurement along line 3 in Fig.~\ref{fig2}(a), we show in Fig.~\ref{fig2}(d) that the emission is absent in the upper band (dashed line) within the first Bz, and in the lower band
(solid line) within the second Bz. This phenomenon arises from destructive interference in the far field emission along certain high symmetry directions. It occurs in lattices with multiple sites per unit cell~\cite{Shirley1995} and has been observed along the $K$-$\Gamma$-$K^{'}$ directions in angle-resolved electron spectroscopy measurements in graphene~\cite{Bostwick2006}.
\begin{figure}[b]
\includegraphics[scale=0.61]{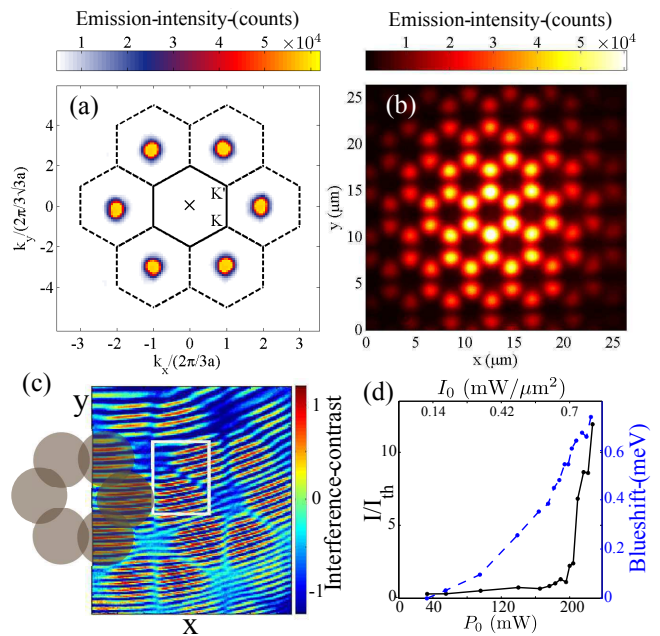}
\caption{(a) Photoluminescence emission in momentum space above the condensation threshold. The black solid/dashed line shows the first/second Bz. (b) Real space image of the condensed state. (c) Interference pattern above condensation threshold. The position of six pillars is underlined with gray disks. (d) Total emitted intensity (black line) and blueshift of the polariton emission at the top of the $\pi^*$ band (blue dashed line) as a function of excitation intensity $I_0$ or power $P_0$.}
\label{fig3}
\end{figure}

By increasing the excitation intensity, we observe polariton condensation, as evidenced by the threshold in the integrated emission intensity [Fig.~\ref{fig3}(d)]. The threshold power is similar to that observed in a planar structure~\cite{Wertz2009}. The low value of the measured emission blueshift, due to interactions between polaritons and uncondensed excitons [see Fig.~\ref{fig3}(d)], certifies that the system remains in the strong coupling regime across the threshold~\cite{Ferrier2011}. Moreover, the emission spectrum collapses into a single emission line, and extended spatial coherence builds up. By monitoring the energy resolved emitted intensity across the condensation threshold, we observe that condensation takes place at the top of the $\pi^*$ band [arrow in Fig.~\ref{fig2}(b), \cite{EPAPS}]. This state is located at the $\Gamma$ point (center of the Bzs) as seen in Fig.~\ref{fig3}(a). The far field destructive interference discussed above results in the absence of emission from the center of the first Bz, marked by a cross in Fig.~\ref{fig3}(a).

The real space emission of the condensate is shown in Fig.~\ref{fig3}(b), covering the same area as the pump spot. The intensity maxima are centered on the pillars as expected for a state arising from the hybridization of $S$ states. We extract its phase structure as follows: we magnify the image of one pillar, and make it interfere with an image of the whole excited region~\cite{Manni2012}. The normalized interference pattern, without energy selection, is shown in Fig.~\ref{fig3}(c) above the condensation threshold. We observe spontaneous coherence over the whole size of the pump beam. At the intersection between two adjacent pillars [white square in Fig.~\ref{fig3}(c)], the fringes are shifted by half a period. Thus there is a $\pi$ phase shift between adjacent pillars, as expected for the antibonding $\pi^*$ band. Note that condensation does not take place in the ground state. This feature arises from the out of equilibrium nature of polaritons in which the steady state is fixed by the interplay between pump, relaxation and decay~\cite{Lai2007, Tanese2013a}. The antibonding mode at the $\Gamma$ point favors condensation due to two features: (\textit{i})~its negative effective mass and
positive interaction energy, (\textit{ii})~its longer lifetime, which stems from the antisymmetric
character of the state~\cite{Aleiner2012}, and from the lower nonradiative recombination rate due to the vanishing polariton density at the constrictions between pillars, where the defect density is larger.

\begin{figure}
\includegraphics[scale=0.29]{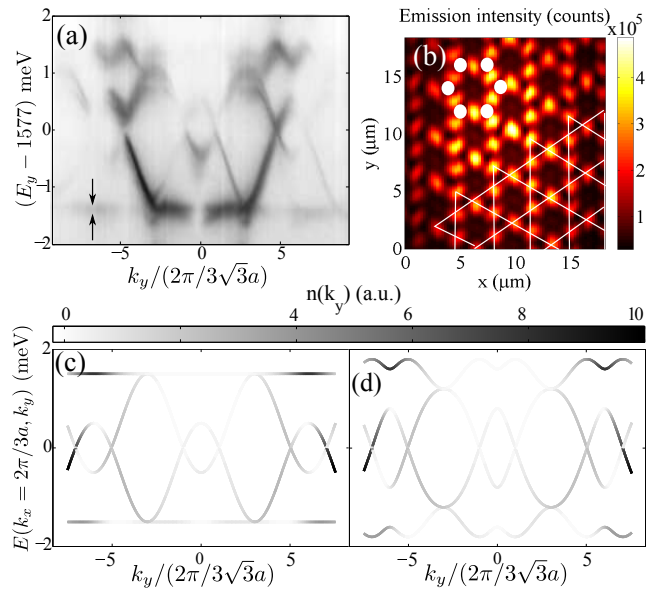}
\caption{(a) Zoom on the four $P$ bands shown in Fig.~\ref{fig1}(d). (b) Real space image integrated in energy over the flatband, marked with two arrows in (a). The centers of six pillars are shown in white disks. The kagome geometry of the emission lobes is underlined with white lines. (c) and (d): calculated energy dispersion at $k_x = -2\pi/3a$ [same direction as in (a)] from Eq.~\ref{TBflat} for $t_\parallel=-1$~meV, $t_\perp=0$~meV (c) and $t_\perp=0.2$~meV (d). For each state, the color scale indicates the
relative brightness of its emission at energy $E$ and wavevector $k_y$,
as predicted by the Fourier transform of the spatial wavefunction.}
\label{fig4}
\end{figure}

We have shown that the two $S$ bands mimic the graphene $\pi$ and $\pi^*$ bands. But the honeycomb lattice contains more than those bands if higher orbital modes are available. In our lattice, the coupling between $P$ modes of the pillars leads to four energy bands which appear above the two $S$ bands, separated by a gap of about 0.7~meV [see Fig.~\ref{fig1}(d)]. The $P$ bands are shown in detail in Fig.~\ref{fig4}(a) revealing that the lowest one is flat. Flatbands are characterized by an infinite effective mass and, consequently, a vanishing kinetic energy. In this situation, one can show that all states are localized without interaction~\cite{Petrescu2012}. Moreover, weak interactions have been predicted to give rise to strongly correlated phases in a
lossless system~\cite{Huber2010, Wu2007}. To understand the origin of the flatbands, one can extend the usual tight-binding treatment to $P$ states with a Hamiltonian of the
form~\cite{Wu2007, Wu2008}
\begin{eqnarray}
\hat{H}=-\sum_{\langle i,j\rangle}
&~&\left[t_\parallel (\hat{\vec{\psi}}_i^\dagger\cdot e_{ij}^{(L)})(
e_{ij}^{(L)\dagger}\cdot \hat{\vec{\psi}}_j)\right.\nonumber \\
&+&
\left.t_\perp (\hat{\vec{\psi}}_i^\dagger\cdot e_{ij}^{(T)})(
e_{ij}^{(T)\dagger}\cdot \hat{\vec{\psi}}_j)
+\textrm{H.c.}\right]
\label{TBflat}
\end{eqnarray}
For each $ij$ link, the $e_{ij}^{(L,T)}$ unit vectors are directed
respectively along and orthogonally to the link direction. In the
Hamiltonian, they serve to extract the projections of the $P$ state
respectively along and orthogonal to the link. The $t_\parallel$
amplitude then describes hopping between $P$ states with main lobes
located along the link, while $t_\perp$ describes the (typically much
weaker) hopping between states with lobes located sideways to the link.
In the limiting case where $t_\perp = 0$~meV and $t_\parallel = -1$~meV, the eigenstates of Eq.~(\ref{TBflat}) give rise to four energy bands plotted in Fig.~\ref{fig4}(c). The two extreme bands are flat, the two intermediate ones are dispersive. For those parameters, this model describes well the lower bands observed in the experiment [Fig.~\ref{fig4}(a)]. However, the higher energy band in Fig.~\ref{fig4}(a) is not flat.
This can be explained by
allowing for a weak hopping also for the \textit{P} states orthogonal to the link. Indeed in the case where $t_\perp = 0.2$~meV and $t_\parallel = -1$~meV, the tight-binding result is plotted in Fig.~\ref{fig4}(d) where the two extreme bands are no longer flat. The band structure reported in Fig.~\ref{fig4}(a) can then be understood assuming that $t_\perp$ increases with the energy, resulting in a flat band ($t_\perp\simeq 0$) at low energy and a dispersive band ($t_\perp\simeq 0.2$~meV) at higher energy. Indeed, the tunneling probability varies exponentially with the barrier height relative to the state, and thus increases strongly for higher energy states. In order to confirm this model, we have performed a numerical simulation of the two-dimensional Schr\"odinger equation which reproduces the observed dispersion~\cite{EPAPS}.
Finally, Fig.~\ref{fig4}(b) shows the flatband real space mode for which intensity maxima sit between the pillars, thus arranged in a kagome geometry.

In summary, we have implemented a system which allows direct optical access to the basic properties of engineered lattices as demonstrated by the direct observation of Dirac cones in a honeycomb geometry. The position, shape and size of each lattice site can be controlled at will during fabrication. Moreover, via resonant excitation of the structure, polariton wave packets can be created with any desired energy and momentum. This configuration has been previously used to evidence polariton flow without scattering and
the hydrodynamic nucleation of vortices and solitons~\cite{Carusotto2013}. It opens the way to study a number of effects in the honeycomb lattice, like Klein tunneling at a potential step~\cite{Allain2011}, the geometrical Berry curvature of the bands~\cite{Ozawa2013} and the topological physics in the presence of synthetic gauge fields~\cite{Rechtsman2012b}. The observation of a bright flat band suggests the possibility of using a resonant pump to selectively inject polaritons into it, and investigate the interplay between frustration, dispersion and interactions in such flatbands~\cite{Wu2007, Huber2010}.

\begin{acknowledgments}
This work was supported by the French RENATECH, the ANR-11-BS10-001 contract "QUANDYDE", the RTRA Triangle de la Physique (Contract "Boseflow1D"), the FP7 ITNs "Clermont4" (235114), the FP7 IRSES "Polaphen" (246912), the POLATOM ESF Network, the Labex Nanosaclay, and the ERC (Honeypol and QGBE).
\end{acknowledgments}

\begin{center}
\textbf{SUPPLEMENTARY MATERIAL}\\
\end{center}

\section{Condensation in the $\pi^{*}$ band: experiments}

\begin{figure}[h!]
\includegraphics[scale = 0.65]{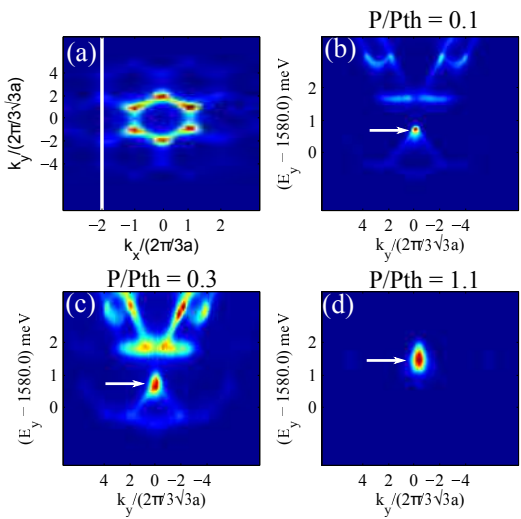}
\caption{Photoluminescence emission in the far field: (a)  at the Dirac points energy at very low pump intensity, (b)  energy dispersion along the white line in (a) at pump power $P/P_{\rm{th}}$ = 0.1,  (c) $P/P_{\rm{th}}$ = 0.3 and (d) $P/P_{\rm{th}}$ =  1.1}
\label{condensationPi}
\end{figure}

In order to prove that the state at which polariton condensation takes place is located at the top of the $\pi^*$ band, as reported in Fig.~3, we show here a detailed power dependence of the emission across the condensation threshold. At low power [Fig.~\ref{condensationPi}(b)], below threshold, all the low energy bands are populated. At the $\Gamma$ point a brighter point is observed showing efficient relaxation towards that state. When we approach the threshold for condensation we observe that particles start to accumulate at the top of the $\pi^*$ band [Fig.~\ref{condensationPi}(c)]. Above threshold it is that particular state the one that becomes macroscopically occupied (Fig.~\ref{condensationPi}(d)). Note that the $\pi$ and $\pi^*$ bands continuously blueshift when increasing the excitation power due to the repulsive interactions between polaritons populating that band and the highly populated exciton reservoir located at the bare exciton energy (about 20~meV above in energy).

\section{Condensation in the $\pi^{*}$ band: simulations}

To simulate polariton condensation in the honeycomb structure we have used a 2D Gross-Pitaevskii equation with additional terms describing the polariton lifetime, spontaneous polariton scattering (noise), stimulated scattering term (included in the form of a saturated gain, accounting for scattering from the reservoir) and kinetic energy relaxation that takes the form of an energy-dependent decay term~\cite{Wertz2012}:

\begin{eqnarray}
i\hbar\frac{\partial \Psi}{\partial t} = &~& -(1-i \Lambda) \frac{\hbar^2}{2m} \Delta \Psi + \alpha \left| \Psi \right| ^{2}\Psi  - \frac{i \hbar}{2\tau}\Psi \nonumber\\
&~& + \Biggl(U(\textbf{r})+U_{R}(n) exp\Biggl(-\frac{(\textbf{r} - \textbf{r$_{0}$})^2}{\sigma^2}\Biggr)\Biggr)\Psi
\nonumber\\
&~& + i\gamma(n) exp\Biggl(-\frac{(\textbf{r} - \textbf{r$_{0}$})^2}{\sigma^2}\Biggr) \Psi+\xi.
\label{cond}
\end{eqnarray}

Here $m$ is the polariton mass, $\Lambda=3\times 10^{-3}$ is the kinetic energy relaxation term, $\alpha=3 E_b a_{b}^{2}$ is the polariton-polariton interaction constant ($E_b =10$~meV is the exciton binding energy and  $a_b=10$~nm is the exciton Bohr radius), $U(\textbf{r})$ is the honeycomb lattice potential (height 20 meV), containing an imaginary part accounting for the shorter lifetime induced by the evanescent part of the modes outside of the pillars. $U(\textbf{r})$ eventually gives rise to the honeycomb dispersion, including the $S$ and $P$ bands. $U_R(n)$ is the potential induced by the reservoir, which we take to be equal to 1~meV for the considered injected polariton density $n$. The reservoir has a Gaussian shape with a width of $45~\mu$m given by the size of the excitation spot. $\tau$ is the polariton lifetime (30 ps), $\gamma(n)$ is the saturated stimulated scattering rate from the reservoir to the condensate, and $\xi$ is the Gaussian noise term with amplitude $10^{-3}\hbar / 2\tau$.

For this set of parameters the simulations reproduce condensation at the $\Gamma$ point on the top of the $\pi^*$ band, as in the experiment. The condensation mechanism in that negative mass state can be understood as follows. First, the reservoir of excitons created by the nonresonant pump creates a repulsive potential for polaritons, which pushes away particles created by spontaneous scattering, preventing the formation of the condensate in the states with positive mass. However, the states with negative mass are on the contrary trapped in this potential, and serve as a seed for stimulated scattering. A second reason for the condensation of polaritons on top of the $\pi^*$ band is that the lifetime of anti-symmetric states is in general longer than that of the symmetric one~\cite{Aleiner2012}. This is due to the fact that the evanescent fraction of the mode outside the pillars is reduced for these modes due to the presence of the zeroes of the wavefunction at all junctions between the pillars, where there is a larger density of non-radiative centers that contribute to the lifetime reduction. This aspect favors the $\Gamma$ point of the $\pi^*$ band with respect to (for example) the $\Gamma$ points of the non-flat $P$ bands, which might also have negative mass, or with respect to the flat band, which possess much shorter lifetimes due to the location of the wavefunction lobes on the junctions between the pillars (see Fig.~4(b) of the main text).

\begin{figure}
\includegraphics[scale = 0.55]{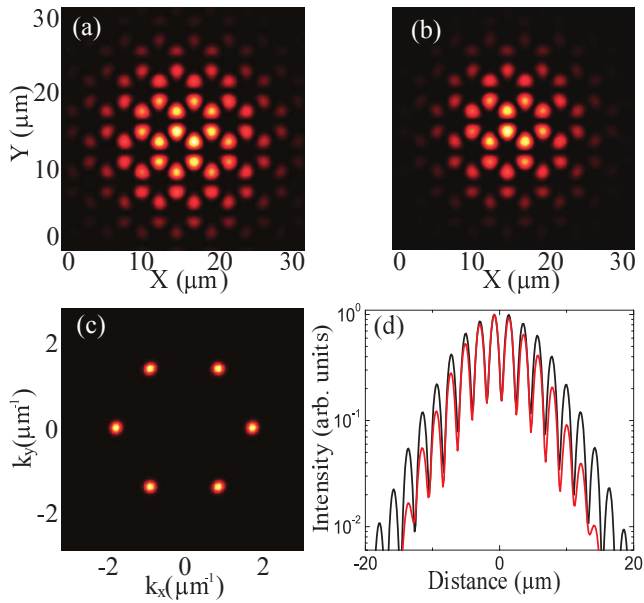}
\caption{Spatial image of the condensate constructed from the simulation of the modified 2D Gross-Pitaevskii equation (Eq.\ref{cond}) with (a) negligible interactions $\alpha \left| \Psi \right| ^{2} \ll U_R(n)$ and (b) significant interactions $\alpha \left| \Psi \right| ^{2} \sim U_R(n)$. (c) Fourier transform of the simulated emission corresponding to (a). (d) Spatial transverse profile passing through the center of the excitation spot extracted from (a) (black) and (b) (red).}
\label{figSimulCond}
\end{figure}

Figure~\ref{figSimulCond}(a) shows the simulated emission from the condensate in the real space and Fig.~\ref{figSimulCond}(c) in the reciprocal space, in the absence of polariton-polariton interactions ($\alpha \left| \Psi \right| ^{2} \ll \frac{ \hbar}{2\tau}, U_R(n)$). The simulation is in quantitative agreement with the experimental observations (Figs.~3(b) and~(a), respectively), including the absence of emission from the $\Gamma$ point in the first Brillouin zone due to interference effects. When varying the poition of the pump spot with respect to the center of the lattice, a very similar spatial and momentum space patterns are obtained.

The spatial extension of the condensate coincides with that of the excitation spot that populates the reservoir. When interactions in the condensate become non-negligible compared to the interactions induced by the reservoir ($\alpha \left| \Psi \right| ^{2} \sim U_R(n)$) we expect the state to evolve into a gap soliton bound to the reservoir, as a consequence of the same mechanisms that have allowed its observation in a 1D periodic lattice for polaritons~\cite{Tanese2013}. The increase of the polariton-polariton interaction term $\alpha \left| \Psi \right| ^{2}$ in the simulation leads to the shrinking of the spatial extension of the emission (see Fig.~\ref{figSimulCond}(b)). Even the smallest interactions within the condensate bring its energy up, further into the gap. The modification of the simulated transverse profile of the condensate corresponding to Fig.~\ref{figSimulCond}(a, b)) is shown in Fig.~\ref{figSimulCond}(d). Limitation in the highest available excitation density in the experiment prevents us from seeing the expected modification in the spatial profile when the condensate evolves into a gap soliton.

\section{$P$ bands: 2D Schr\"odinger equation simulation}

In order to confirm the phenomenological model used to describe the results reported in Fig.~4, in which we assume that the tunnelling probability is energy dependent, we have performed a 2D Schr\"odinger equation simulation for polaritons in the low density limit. Since the $S$ and $P$ bands are located close to the bottom of the lower polariton branch, we use the effective mass approximation:

\begin{eqnarray}
i\hbar\partial_t \Psi = &~& -\frac{\hbar^2}{2m}\Delta \Psi + \left(U - \frac{i\hbar}{2\tau}\right)\Psi \nonumber\\
&~&+P_0e^{-\frac{(t-t_0)^2}{\tau_0^2}}e^{-\frac{(\bf{r}-\bf{r_0})^2}{\sigma^2}}e^{-i\omega t}. 
\label{schrod}
\end{eqnarray}

Here $m$ is the polariton mass, $\tau = 30~$ps is the polariton lifetime, and $U$ is the external potential describing the etched honeycomb lattice. In our simulation we use a rectangular sample made out of coupled micropillars of round geometry and same dimension as in the experiment, arranged in a lattice with 16 by 16 unit cells. The height of the polariton confining potential in the micropillars was taken 20~meV. The last term of the equation simulates a pulsed probe that will excite the different eigenstates of the Schr\"odinger equation, thus allowing their visualization. $P_0$ is the amplitude of the probe, arriving at the sample at $t_0$, $\tau_0 = 0.2$~ps is the pulse duration, $\sigma = 0.7~\mu$m the spot size. Using a short pulse and a small spot allows exciting several bands of the dispersion at the same time. $\bf{r}_0$ is the pump location (center of the sample, which does not correspond to the center of a particular pillar) and $\omega$ is the pump central frequency, centered 4~meV above the bottom of the lower polariton branch to mainly excite the $P$ band multiplet. Let us note that the probe pulse excites different parts of the dispersion with different efficiency, depending on their symmetry. 

\begin{figure}
\includegraphics[scale = 0.57]{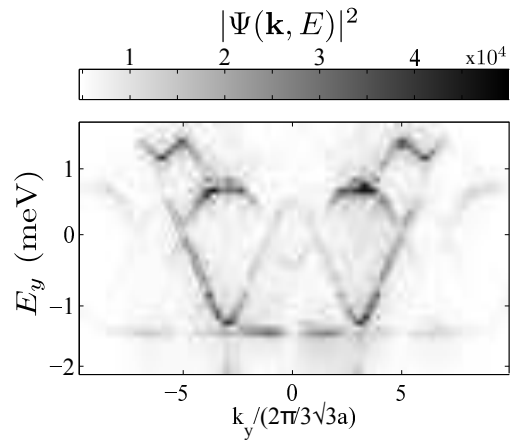}
\caption{Simulation of $|\Psi(\textbf{k}, E)|^2$ along the same momentum space direction as Fig.~4(a) in the main text, based on the solution of Eq.~\ref{schrod}.}
\label{figepaps}
\end{figure}

The Schr\"odinger equation is then integrated over time for 100~ps with a spatial grid 512x512 (the size of the grid is $80\times 80\mu$m) using a NVIDIA graphic card. The solution of the equation $\Psi(\textbf{r}, t)$ is then Fourier-transformed over time and space to obtain the dispersion $|\Psi(\textbf{k}, E)|^2$. The result is shown in Fig.~\ref{figepaps} along the same momentum-space direction as in Fig.~4(a) of the main text. The simulation is in excellent quantitative agreement with the experimental observation: the lowest $P$ band is indeed flat, while the upper band is dispersive. The full 2D model reproduces this behavior correctly, because it automatically takes into account the exponential increase with energy of the tunneling rate of the $P$ states of the individual pillars, as explained in the main text and illustrated by the tight-binding model calculations.

\end{document}